\documentclass[12pt]{article}
\pdfoutput=1

\usepackage[bulletsep]{collref}
\usepackage{amssymb,graphicx}
\usepackage[intlimits]{amsmath}
\usepackage{pst-all}
\usepackage{bbm}
\usepackage[small]{subfigure}

\usepackage{MnSymbol}

\makeatletter \@addtoreset{equation}{section} \makeatother

\makeatletter
\let\old@startsection=\@startsection
\let\oldl@section=\l@section
\renewcommand{\@startsection}[6]{\old@startsection{#1}{#2}{#3}{#4}{#5}{#6\mathversion{bold}}}
\renewcommand{\l@section}[2]{\oldl@section{\mathversion{bold}#1}{#2}}
\makeatother

\makeatletter
\let\old@makecaption=\@makecaption
\def\@makecaption{\small\old@makecaption}
\makeatother

\begin{document}

\thispagestyle{empty}

\begin{flushright}\footnotesize
\texttt{NORDITA-2013-1} \\
\texttt{UUITP-01/13}\\
\texttt{UWO-TH-13/1}
\vspace{0.6cm}
\end{flushright}

\renewcommand{\thefootnote}{\fnsymbol{footnote}}
\setcounter{footnote}{0}

\def\caln{\mathcal{N}}

\begin{center}
{\Large\textbf{\mathversion{bold} Rigorous Test of Non-conformal Holography:  
\\
Wilson Loops in $\mathcal{N}=2^*$ Theory
}
\par}

\vspace{0.8cm}

\textrm{Alex~Buchel$^{1,2}$, Jorge~G.~Russo$^{3,4}$  and Konstantin~Zarembo$^{5,6}$\footnote{Also at ITEP, Moscow, Russia}}
\vspace{4mm}

\textit{${}^1$ Perimeter Institute for Theoretical Physics\\
 Waterloo, Ontario N2J 2W9, Canada}\\
\textit{${}^2$ Department of Applied Mathematics,
 University of Western Ontario\\
 London, Ontario N6A 5B7, Canada}\\
\textit{${}^3$ Instituci\'o Catalana de Recerca i Estudis Avan\c cats (ICREA), \\
Pg. Lluis Companys, 23, 08010 Barcelona, Spain}\\
\textit{${}^4$ ECM Department, Facultat de F\'\i sica, Universitat de Barcelona, \\
Marti i Franqu\`es, 1, 08028 Barcelona, Spain}\\
\textit{${}^5$Nordita, KTH Royal Institute of Technology and Stockholm University,
Roslagstullsbacken 23, SE-106 91 Stockholm, Sweden}\\
\textit{${}^6$Department of Physics and Astronomy, Uppsala University\\
SE-751 08 Uppsala, Sweden}\\
\vspace{0.2cm}
\texttt{abuchel@perimeterinstitute.ca, russo.jg@gmail.com, zarembo@nordita.org}

\vspace{3mm}


\par\vspace{0.4cm}

\textbf{Abstract} \vspace{3mm}

\begin{minipage}{13cm}
$\caln=2^*$ gauge theory in four space-time dimensions arises as a
deformation of the parent $\caln=4$ supersymmetric Yang-Mill theory
under which its hypermultiplet acquires a mass.  This theory, on the one hand, has a known supergravity dual and on the other hand is amenable to localization. We explain, using localization, a dynamical selection of the supergravity
Coulomb branch vacuum at large $N$. We also demonstrate that large Wilson loops obey perimeter law, again finding exact match between direct field-theory calculations and string theory. Additionally, 
we compute free energy at large 't Hooft coupling. 
\end{minipage}

\end{center}

\vspace{0.5cm}


\newpage
\setcounter{page}{1}
\renewcommand{\thefootnote}{\arabic{footnote}}
\setcounter{footnote}{0}

\section{Introduction}

Holographic duality is a powerful tool to study strongly-coupled systems, but direct comparison of its predictions with underlying field theory requires doing non-perturbative calculations,  in general a very difficult task. Remarkably, in  $\mathcal{N}=4$ super-Yang-Mills (SYM) theory many such calculations have been done, leading to spectacular confirmations of the AdS/CFT duality. Methods used there look very model-dependent, as they typically rely on the high degree of supersymmetry of $\mathcal{N}=4$ SYM, conformal invariance and integrability. Nevertheless, similar methods can be applied to a deformation of $\mathcal{N}=4$ SYM that breaks both conformal invariance and half of the supersymmetry. 

The theory we will be dealing with is obtained from $\mathcal{N}=4$ SYM by giving common masses to four scalars out of six and to half of the fermions. Such a deformation breaks half of the original supersymmetries and is known as $\mathcal{N}=2^*$ theory. The  gravity dual of this theory is an explicitly known solution of ten-dimensional supergravity \cite{Pilch:2000ue}. On the field-theory side, the main technical advance is localization
 \cite{Pestun:2007rz}  allowing one to do computations at any coupling. We will exploit these two facts to test holographic duality in this non-conformal setting.

The $SU(N)$ gauge symmetry of $\mathcal{N}=2^*$ theory is spontaneously broken to $U(1)^{N-1}$ by the vev of one of the massless scalars:
\begin{equation}\label{eigenv}
 \left\langle \Phi \right\rangle=\mathop{\mathrm{diag}}\left(a_1,\ldots ,a_N\right).
\end{equation}
The theory thus is not confining and always remains in the Coulomb phase. 
Its  Seiberg-Witten curve is constructed in \cite{Donagi:1995cf}.
At large $N$ the eigenvalues form a continuous distribution conveniently characterized by the density
\begin{equation}
 \rho (x)=\frac{1}{N}\,\sum_{i=1}^{N}\delta \left(x-a_i\right).
\end{equation}

The Pilch-Warner (PW) supergravity background \cite{Pilch:2000ue} corresponds to a very particular eigenvalue density, identified in \cite{Buchel:2000cn}. In the  'Pilch-Warner' Coulomb branch vacuum, the eigenvalues form a semi-circular distribution: 
\begin{equation}\label{Wigner}
 \rho (x)=\frac{2}{\pi \mu ^2}\,\sqrt{\mu ^2-x^2}.
\end{equation}
The width of the distribution is proportional to the mass-deformation parameter $M$ (as expected on dimensional grounds) and the square root of the 't~Hooft coupling $\lambda =g_{YM}^2N$:
\begin{equation}\label{sugrawidth}
 \mu \equiv \mu_{\rm SG} =\frac{\sqrt{\lambda }M}{2\pi }\,.
\end{equation}
This result is obtained within the classical supergravity approximation, and is valid when $\lambda \gg 1$.  

The width of the eigenvalue distribution sets the scale of gauge symmetry breaking. The symmetry-breaking vevs and the masses of W-bosons are on average proportional to $\mu $, which at strong coupling is parametrically bigger than the soft mass $M$ that triggered the symmetry breaking.
However, some of the W-bosons, that connect nearest neighbours in the eigenvalue distribution, are very light, with masses of order $\mu /N$.

We will consider $\mathcal{N}=2^*$ theory on the four-sphere of radius $R$, which here we regard merely as an IR regularization, and will use localization results \cite{Pestun:2007rz} to analyze the theory at large $N$ in the regime of the strong 't~Hooft coupling. We will show that the semicircular eigenvalue distribution (\ref{Wigner}) arises dynamically as the large-$N$ master field, with the width that precisely matches the supergravity prediction (\ref{sugrawidth}). We will then compare the expectation value of the circular Wilson loop with the minimal area law in the PW geometry.

\section{Strong coupling from localization}

The partition function of $\mathcal{N}=2^*$ theory on $S^4$ can be calculated using localization \cite{Pestun:2007rz}.  The partition function reduces to an $(N-1)$-dimensional integral of an effective matrix model:
\begin{equation}\label{main_partition}
 Z=\int d^{N-1} a \,\prod_{i<j} \frac{(a_i-a_j)^2H^2(a_i-a_j)}{H(a_i-a_j-M)H(a_i-a_j+M)}  \,{\rm e}\,^{-\frac{8\pi^2N}{\lambda }\sum_{j}a_j^2} \left|\mathcal{Z}_{\rm inst}\right|^2.
\end{equation}
The integration is over the eigenvalues of the adjoint scalar from the vector multiplet, the same as in eq.~(\ref{eigenv}). All other degrees of freedom have been integrated out. Their contribution is expressed in terms of a single function
\begin{equation}\label{H(x)}
 H(x)\equiv \prod_{n=1}^\infty \left(1+\frac{x^2}{n^2}\right)^n e^{-\frac{x^2}{n}}.
\end{equation}
To avoid notational clutter we express all dimensional quantities in the units set by the radius of $S^4$. The dependence on $R$ can be re-instated by rescaling $a_i\rightarrow a_iR$, $M\rightarrow MR$, which we will do at the very end of the calculation.

The instanton partition function $\mathcal{Z}_{\rm inst}$ is known \cite{Nekrasov:2002qd,Nekrasov:2003rj,Okuda:2010ke}, but is not important for our analysis, since in the large-$N$ limit instantons are exponentially suppressed. We thus set $\mathcal{Z}_{\rm inst}$ to one. 

Observables that can be computed using localization include the free energy:
\begin{equation}
 F=-\ln Z
\end{equation}
and the Wilson loop for the contour running along the big circle of $S^4$:
\begin{equation}\label{WC}
 W(C_{\rm circle})=\left\langle 
 \frac{1}{N}\mathop{\mathrm{P}}\exp\left[\oint 
 ds\,\left(i\dot{x}^\mu A_\mu +|\dot{x}|\Phi \right)
 \right]
 \right\rangle,
\end{equation}
where $\Phi $ is the field that takes on the vacuum expectation value (\ref{eigenv}). Supersymmetry allows linear combinations of the two massless scalars to appear in the exponent, but only this particular operator commutes with the supercharge used in localization as the BRST operator. Therefore localization can compute the Wilson loop vev only with this specific scalar coupling. The Wilson loop vev maps to an exponential average in the matrix model:
\begin{equation}
\label{Wils}
 W\left(C_{\rm circle}\right)=\left\langle \frac{1}{N}\sum_{j}
 \,{\rm e}\,^{2\pi a_j}
 \right\rangle.
\end{equation}

The localization partition function for $\mathcal{N}=4$ SYM ($M=0$) is a simple Gaussian model, easily solvable at large $N$. The Wilson loop vev can then be computed at any coupling. At strong coupling, it grows as $\,{\rm e}\,^{\sqrt{\lambda }}$, in precise agreement with the area law for classical strings in $AdS_5\times S^5$ \cite{Erickson:2000af,Drukker:2000rr}. The study of the matrix integral (\ref{main_partition}) at large-$N$ was initiated in \cite{Russo:2012kj,Russo:2012ay}, following earlier work on similar models \cite{Rey:2010ry,Passerini:2011fe,Bourgine:2011ie,Fraser:2011qa}. 

The integral (\ref{main_partition}) is of the saddle-point type in the large-$N$ limit. The saddle-point equations are
\begin{equation}
 \sum_{k\neq j}^{}\left(
 \frac{1}{a_j-a_k}-K(a_j-a_k)+\frac{1}{2}\,K(a_j-a_k+M)+\frac{1}{2}\,K(a_j-a_k-M)
 \right)=\frac{8\pi ^2}{\lambda }\,a_j,
\end{equation}
where $K(x)$ is the logarithmic derivative of $H(x)$:
\begin{equation}
 K(x)=-\frac{H'(x)}{H}=x\left(\psi \left(1+ix\right)+\psi \left(1-ix\right)-2\psi (1)\right) .
\end{equation}

The continuum limit converts the saddle-point equation into a singular integral equation for the eigenvalue density:
\begin{equation}
 \label{nnstar}
\strokedint_{-\mu}^\mu dy\, \rho(y) \left(\frac{1}{x-y} -K(x-y)+{1\over 2}\,K(x-y+M)+\frac{1}{2}\,K(x-y-M)\right)= \frac{8\pi^2}{\lambda}\, x.
\end{equation}
The density is defined on the interval $(-\mu ,\mu )$, and the equation also holds only on this interval. This equation is difficult to solve in general, but its solution at strong coupling can be obtained by a slight modification of the small-$M$ limit considered in \cite{Russo:2012kj}.

The basic assumption, that we will check {\it a posteriori}\footnote{We have made an extensive set of numerical tests which confirm the validity of this assumption.}, is that at $\lambda \rightarrow \infty $ the symmetry-breaking scale is much larger than the hypermultiplet mass, as suggested by the supergravity analysis:  $\mu \gg M $. This justifies an approximation
\begin{equation}
 {1\over 2}\,K(x+M)+\frac{1}{2}\,K(x-M)-K(x)\approx 
 \frac{1}{2}\,K''(x)M^2\approx \frac{M^2}{x}\,,
\end{equation}
after which the integral equation (\ref{nnstar}) greatly simplifies:
\begin{equation}
 \strokedint_{-\mu}^\mu dy\, \rho(y) \,\frac{1+M^2}{x-y}= \frac{8\pi^2}{\lambda}\, x,
\end{equation}
and is solved by Wigner's semi-circular distribution (\ref{Wigner}) with
\begin{equation}\label{widthofWigner}
 \mu =\frac{\sqrt{\lambda \left(1+M^2\right)}}{2\pi }\,.
\end{equation}

The solution is written in the $R=1$ units. Recovering dependence on the radius, we find:
\begin{equation}
  \mu =\frac{\sqrt{\lambda \left(M^2+\frac{1}{R^2}\right)}}{2\pi }\,.
\end{equation}
This result exactly agrees with the supergravity prediction (\ref{sugrawidth}) in the decompactification limit $R\rightarrow \infty $.

We now turn to the Wilson loop vev:
\begin{equation}
  W\left(C_{\rm circle}\right)=\int_{-\mu }^{\mu }dx\,\rho (x)
  \,{\rm e}\,^{2\pi x}.
\end{equation}
The integral here is highly peaked at the largest eigenvalue, such that
\begin{equation}\label{finitecircle}
 \ln W\left(C_{\rm circle}\right)=\sqrt{\lambda \left(1+M^2\right)}
 + O\left(\lambda ^0\right).
\end{equation}
Again, $M$ should be understood as $MR$. Taking the limit of $R\rightarrow \infty $, we find that the Wilson loop satisfies the perimeter law:
\begin{equation}\label{matrixarealaw}
  \ln W\left(C_{\rm circle}\right)=\sqrt{\lambda }\ MR\qquad \left(R\rightarrow \infty \right).
\end{equation}

We can also compute the free energy, by first differentiating in $M$:
\begin{eqnarray}
 \frac{\partial F}{\partial M}&=&\frac{N^2}{2}
 \int_{}^{}dxdy\,\rho (x)\rho (y)
 \left(K(x-y-M)-K(x-y+M)\right)
\nonumber \\
&\approx& 
 -N^2M\int_{}^{}dxdy\,\rho (x)\rho (y)K'(x-y)
\nonumber \\
&\approx& 
 -2N^2M\int_{}^{}dxdy\,\rho (x)\rho (y)
 \left(\ln|x-y|+1+\gamma \right),
\end{eqnarray}
 where $\gamma =-\psi (1)$ is the Euler's constant.
Plugging in the semicircular distribution (\ref{Wigner}) with the width (\ref{widthofWigner}), we get:
 \begin{equation}
 \frac{\partial F}{\partial M}=-N^2M\ln\frac{\lambda \left(1+M^2\right)\,{\rm e}\,^{2\gamma +\frac{3}{2}}}{16\pi ^2}\,.
\end{equation}
Choosing the normalization constant such that at $M=0$ the $\mathcal{N}=4$ SYM result is recovered, we find:
\begin{equation}\label{freeE}
 F=-\frac{N^2}{2}\,\left(1+M^2\right)\ln\frac{\lambda \left(1+M^2\right)\,{\rm e}\,^{2\gamma +\frac{1}{2}}}{16\pi ^2}\,.
\end{equation}

For a large sphere we find, with a logarithmic accuracy:
\begin{equation}
 F=-N^2M^2R^2\ln MR\qquad (R\rightarrow \infty ).
\end{equation}

\section{Strong coupling from supergravity}

The supergravity dual of the $\mathcal{N}=2^*$ theory is a warped product of a deformed $AdS_5$ and a deformed five-sphere \cite{Pilch:2000ue}. The warp factor, the dilaton, the form fields and the metric all depend on the AdS radius and on the azimuthal angle on $S^5$.   The deformed five-sphere is foliated by elongated three-spheres, whose  $SU(2)\times U(1)$ isometry realizes geometrically the R-symmetry of the dual field theory.

We will study string solutions that are dual to Wilson loops in SYM. The string then ends on a given contour on the boundary and
sits at one point on the internal manifold. This point is determined by the scalar couplings of the Wilson loop. Locally supersymmetric Wilson loops, amenable to  localization, couple to the symmetry-breaking scalar in the vector multiplet. The coupling to this scalar geometrically corresponds to a locus where the R-symmetry three-sphere  shrinks to a point ($\theta =\pi /2$, in the notations of \cite{Pilch:2000ue,Buchel:2000cn}). On this locus, the AdS metric takes the following form:
\begin{equation}\label{Einmetric}
 ds^2 =\frac{\sqrt{c}\rho ^6}{c^2-1}\,M^2dx_\mu ^2+\frac{\sqrt{c}}{\rho ^6\left(c^2-1\right)^2}\,dc^2,
\end{equation}
where $c$ is the radial coordinate and 
\begin{equation}\label{rho6}
 \rho ^6=c+\frac{c^2-1}{2}\,\ln\frac{c-1}{c+1}\,.
\end{equation}
The boundary is at $c=1$. Near the boundary, the geometry asymptotes to $AdS_5\times S^5$ of unit radius (in this normalization, the radius of AdS appears in the string action in combination with $\alpha '$ which together make the square root of the 't~Hooft coupling). The relationship of $c$ with the standard radial coordinate of $AdS_5$, $z$,  is
\begin{equation}\label{asymptot}
 c=1+\frac{z^2M^2}{2}+O(z^4).
\end{equation}

The dilaton also depends on the polar angle $\varphi $, that determines orientation of the scalar coupling of the Wilson loop with respect with respect to symmetry-breaking vev. For the reasons explained after eq.~(\ref{WC}), we should set this angle to zero.
The dilaton has a very simple profile at $\theta =\pi /2$ and $\varphi=0$:
\begin{equation}
 \,{\rm e}\,^{-\phi }=c.
\end{equation}

In addition, the background contains  a non-trivial B-field and Ramond-Ramond form fields. The Ramond-Ramond fields do not couple directly to the classical string worldsheet, and the two-form potential of the B-field has all its indices on the sphere. The string that moves only in AdS and is shrunk to a point on the internal manifold will not interact with the B-field either. For this reason we do not display explicit formulas for the form fields here. We also omit  a rather complicated metric on the internal manifold. The  fully explicit supergravity solution can be found in \cite{Pilch:2000ue}.

The metric (\ref{Einmetric}) is written in the Einstein frame and is normalized such that the string action is given by 
\begin{equation}
 S=\frac{\sqrt{\lambda }}{2\pi }\int_{}^{}d^2\sigma \,\,{\rm e}\,^{\frac{\phi }{2}}
 \sqrt{\det_{ab}G_{MN}\partial _aX^M\partial _bX^N}.
\end{equation}
An expectation value of the Wilson loop, at strong coupling, is determined by the renormalized on-shell action of the string that ends on the contour $C$ at the boundary of AdS \cite{Maldacena:1998im,Rey:1998ik}:
\begin{equation}\label{logW}
 \ln W(C)=-S_{\rm ren}(C).
\end{equation}
Renormalization consists in integrating from the cutoff surface at $c_\epsilon =1+\epsilon ^2M^2/2$ and subtracting the perimeter divergence:
\begin{equation}\label{reg}
 S_{\rm ren}=S-\frac{\sqrt{\lambda }L(C)}{2\pi \epsilon }\,.
\end{equation}
Here we used the relationship (\ref{asymptot})  between $c$  and the radial coordinate in $AdS_5$ in the asymptotic region  near the boundary. Justification of this regularization prescription can be found in \cite{Drukker:1999zq}.

Consider now a segment of a straight line of length $L\gg{\rm anything~else}$. The minimal surface in this case is just a vertical two-dimensional wall with the induced metric
\begin{equation}
 ds_{2d}^2=\frac{\sqrt{c}\rho ^6}{c^2-1}\,M^2dx ^2+\frac{\sqrt{c}}{\rho ^6\left(c^2-1\right)^2}\,dc^2,
\end{equation}
whose area can be readily calculated:
\begin{equation}
 S=\frac{\sqrt{\lambda }ML}{2\pi }\int_{1+\frac{\epsilon ^2M^2}{2}}^{\infty }
 \frac{dc}{\left(c^2-1\right)^{\frac{3}{2}}}
 =\frac{\sqrt{\lambda }L}{2\pi \epsilon }-\frac{\sqrt{\lambda }LM}{2\pi}
 +O\left(\epsilon \right).
\end{equation}
It is interesting to notice that the complicated factor $\rho ^6$ drops out from this calculation.
From (\ref{logW}) and (\ref{reg}) we then find for the expectation value of the straight line:
\begin{equation}\label{perlaw}
 \ln W(C_{\rm line})=\frac{\sqrt{\lambda }ML}{2\pi}\,.
\end{equation}

It is not difficult to see that the same result holds for any sufficiently large contour. More precisely, for any contour whose radius of curvature is large compared to $1/M$. The minimal surface for an arbitrary contour starts off vertically at the boundary, with deviations from verticality proportional to the local curvature of the contour. If such deviations are small, the minimal surface will remain vertical not only near the boundary but all the way through the transition region with $c\sim 1$, and will close off only in the near-enhan\c{c}on region of $c\gg 1$, where the string-frame metric behaves as 
\begin{equation}\label{nearen}
 ds^2_{\rm s.f.}\sim \frac{1}{c^3}\left(\frac{2}{3}\,M^2dx_\mu ^2+\frac{3}{2}\,dc^2\right).
\end{equation}
The line element there is rather small, so the biggest contribution to the area comes from the transition region of $c\sim 1$, where the minimal surface is a vertical wall. We make this argument more precise in the appendix by considering circular Wilson loop in more detail.

We thus conclude that large Wilson loops at strong coupling loops obey the perimeter law (\ref{perlaw}) (in other words, the area of the minimal surface in the PW geometry is proportional to the perimeter of its boundary).
This perfectly agree with the localization result (\ref{matrixarealaw}) for a circular loop of a very big radius, including the coefficient.

\section{Conclusions}

We have shown that the vacuum structure of $\mathcal{N}=2^*$ (the distribution of eigenvalues of the symmetry-breaking vev) and the expectation value of the circular Wilson loop computed directly from the field-theory path integral perfectly agree with supergravity predictions based on the PW geometry. It is interesting to mention in this respect that semicircular distribution of eigenvalues arises in a very general class of supergravity backgrounds \cite{Carlisle:2003nd}. It would be interesting to understand if the field-theory results (effective reduction to a Gaussian model at strong coupling) also hold in a more general setting than $\mathcal{N}=2^*$ theory. 

The supergravity results are recovered in the decompactification limit, when the radius of the four-sphere $R$ goes to infinity. However, our main results (\ref{widthofWigner}), (\ref{finitecircle}) and (\ref{freeE}) hold on the four-sphere of any radius. The dependence on the radius is not very complicated, and it would be interesting to reproduce it from supergravity. For that one needs to find the (1/2 supersymmetric) supergravity solution whose boundary is $S^4$ rather than $\mathbbm{R}^4$ as in the PW geometry. Perhaps this can be done along the lines of \cite{Buchel:2003qm}.

\subsection*{Acknowledgments}
We would like to thank D.~Young for discussions.
K.Z. is grateful to ICTP-SAIFR, S\~{a}o Paulo, and IIP, Natal for kind hospitality during the course of this work.
The work of K.Z. was supported in part by  the RFFI grant 10-02-01315, and in part
by the Ministry of Education and Science of the Russian Federation
under contract 14.740.11.0347. JR acknowledges support from project FPA 2010-20807.
AB gratefully
acknowledges support by an NSERC Discovery grant.
Research at Perimeter
Institute is supported through Industry Canada and by the Province of Ontario through the Ministry of Research $\&$ Innovation.

\appendix

\section{Circular Wilson loop}

In this appendix we study circular Wilson of radius $a$ in the PW geometry.  The string worldsheet is a surface of revolution parameterized by $r=r(c)$, where $r$ is the radial coordinate on the boundary. The string action is then
\begin{equation}\label{Sactioncirc}
 S=\sqrt{\lambda }\int_{}^{}
 dc\,\,
 \frac{r\rho ^6}{c^2-1}\sqrt{\acute{r}^2+\frac{1}{\rho ^{12}\left(c^2-1\right)}},
\end{equation}
where $\rho ^6$ is given by (\ref{rho6}), and we have set $M=1$ for notational simplicity.

The equation of motion for $r(c)$
\begin{equation}\label{minsurf}
 \left[\frac{r\acute{r}\rho ^6}{\left(c^2-1\right)\sqrt{\acute{r}^2+\frac{1}{\rho ^{12}\left(c^2-1\right)}}}\right]'
 =
 \frac{\rho ^6}{c^2-1}\sqrt{\acute{r}^2+\frac{1}{\rho ^{12}\left(c^2-1\right)}}
\end{equation}
must be supplemented by boundary conditions, one of which is $r(1)=a$ and another one is the condition that the worldsheet closes in the bulk.

The solution for a small Wilson loop with $a\ll 1$ does not go far away from the boundary and coincides with the minimal surface in $AdS_5$: $r^2=a^2-2(c-1)$  \cite{Drukker:1999zq,Berenstein:1998ij}. The renormalized area of this solution is $S_{\rm ren}=-\sqrt{\lambda }$. 

We shall consider the opposite limit of large radius: $a\gg 1$. The minimal surface then extends further into the bulk. In fact, the largest portion of the surface lies in the near-enhan\c{c}on region with $c\gg 1$. Taking into account that $\rho ^6\simeq 2/(3c)$ there, we find that the solution takes on a scaling form:
\begin{equation}\label{scalingform}
 r(c)=a\,f\left(\frac{3c}{2a}\right)\qquad \left(c\gg 1\right),
\end{equation}
 where $f(x)$ is a universal function that satisfies the equation
\begin{equation}
 \left(\frac{f\acute{f}}{x^3\sqrt{\acute{f}^2+1}}\right)'
 =
 \frac{\sqrt{\acute{f}^2+1}}{x^3}\,,
\end{equation}
 which is nothing but the minimal-surface equation in the asymptotic, near-enhan\c{c}on geometry (\ref{nearen}).  It should be supplemented with the boundary conditions
\begin{equation}
 f(0)=1
\end{equation}
and 
\begin{equation}
 \acute{f}(x_0)=\infty ,\qquad f(x_0)=0.
\end{equation}
The second condition makes sure that the surface smoothly closes in the bulk. 

\begin{figure}[t]
\begin{center}
 \centerline{\includegraphics[width=8cm]{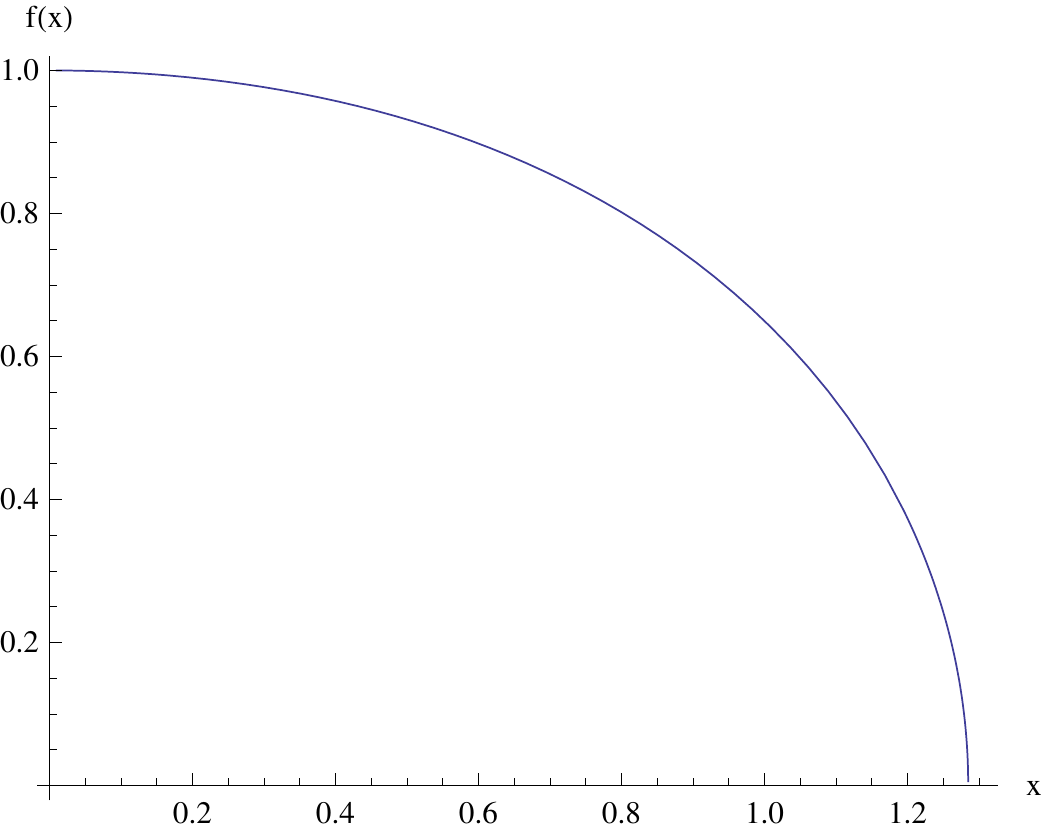}}
\caption{\label{ffig}\small The scaling function $f(x)$. For the endpoint we get $x_0=1.285$. The minimal surface thus extend up to $c_{\rm max}=0.857a$.}
\end{center}
\end{figure}

We were unable to find an analytic form of the scaling function, but numerically it is easy to compute. The result is shown in fig.~\ref{ffig}.
It is also possible to expand $f(x)$ at small $x$:
\begin{equation}\label{fatsmallx}
 f(x)=1-\frac{x^2}{4}+\frac{x^4}{32}\left(\ln x-K\right)
 +\frac{x^6}{96}\left(\ln x-K+\frac{31}{48}\right)+O(x^8\ln^2x).
\end{equation}
The first few coefficients are fixed locally, by the boundary condition at $x=0$, while the constant $K$ is only determined by the boundary condition at $x=x_0$. Numerically, $K=2.200$.

Although the scaling solution covers the largest portion of the minimal surface, it contributes only a small fraction into the total area, because of the rapid decrease of the metric with $c$. Plugging the scaling solution (\ref{scalingform}) into the action (\ref{Sactioncirc}), we find that the area scales as $1/a$:
\begin{equation}\label{SI}
 S_I=\frac{9\sqrt{\lambda }}{4a}\int_{\xi }^{x_0}\frac{dx}{x^3}\,f\sqrt{\acute{f}^2+1}\,.
\end{equation}
The lower cutoff $\xi \ll 1$ should be regarded as a matching scale below which the scaling solution is no longer accurate. The integral actually diverges on the lower bound. Using the expansion (\ref{fatsmallx}), we find:
\begin{equation}\label{largeas}
 S_I=\frac{9\sqrt{\lambda }}{4a}\left(\frac{1}{2\xi ^2}+\frac{\ln \xi }{8}+{\rm finite}\right),
\end{equation}
where 'finite' denotes a cutoff-independent part that stays finite in the limit $\xi \rightarrow 0$.

We can assume that the matching scale is such that $\xi a\gg 1$. Then we can use the scaling solution for $c\gtrsim \xi a$. For $c\lesssim \xi a$ (where $c\sim 1$), the solution can be expanded in power series in $1/a$:
\begin{equation}\label{rin1/R}
 r(c)=a-\frac{u(c)}{a}+\ldots \qquad (c\sim 1).
\end{equation}
Substituting this expansion into the equation of motion (\ref{minsurf}), we find the particular solution
\begin{equation}
 u(c)=\int_{1}^{c}db\,\,\frac{b-\sqrt{b^2-1}}{\left(b+\frac{b^2-1}{2}\,\ln\frac{b-1}{b+1}\right)^2}\,,
\end{equation}
where we have used the explicit form of $\rho ^6$ from (\ref{rho6}). Near the boundary,
\begin{equation}\label{usmall}
 u(c)=c-1+O\left((c-1)^{\frac{3}{2}}\right)\qquad \left(c\rightarrow 1^+\right).
\end{equation}
At large $c$, $u$ grows quadratically:
\begin{equation}\label{ularge}
 u(c)\simeq \frac{9c^2}{16}\qquad \left(c\rightarrow \infty \right).
\end{equation}
This asymptotics, taking into account (\ref{rin1/R}) and (\ref{scalingform}), agrees with the first term in the expansion (\ref{fatsmallx}) in the matching region $c\sim \xi a$.

As discussed in the main text, the solution in the transition region is responsible for the most part of the string area. Plugging (\ref{rin1/R}) into the action (\ref{Sactioncirc}) and expanding in $1/a$, we get:
\begin{equation}\label{SII}
 S_{II}=\sqrt{\lambda }a\int_{c_\varepsilon }^{\frac{2\xi a}{3}}\frac{dc}{\left(c^2-1\right)^{\frac{3}{2}}}
 -\frac{\sqrt{\lambda }}{a}
 \int_{1}^{\frac{2\xi a}{3}}
dc\,
\left[
\frac{u}{\left(c^2-1\right)^{\frac{3}{2}}}
-\frac{\acute{u}}{2}\left(\frac{c}{\sqrt{c^2-1}}-1\right)
\right], 
\end{equation} 
As in (\ref{SI}), we have chosen $c=2\xi a/3$  as the matching scale. In
other words, we use solution (\ref{rin1/R}) for $c<2\xi a/3$ and solution
(\ref{scalingform}) for $c>2\xi a/3$. Using the asymptoticae
(\ref{usmall}) and (\ref{ularge}), and subtracting the perimeter
divergence according to (\ref{reg}), we find that
\begin{equation}\label{smallas}
 S_{II,{\rm ren}}=-\sqrt{\lambda }a-\frac{\sqrt{\lambda }}{a}
 \left(\frac{9}{8\xi ^2}+\frac{9}{32}\ln\xi a+{\rm finite}\right).
\end{equation}
The dependence on the matching scale vanishes in the total action that combines (\ref{largeas}) and (\ref{smallas}), and we find, with logarithmic accuracy:
\begin{equation}
 S_{\rm ren}=-\sqrt{\lambda }a-\frac{9\sqrt{\lambda }\ln a}{32a}\,.
\end{equation}
Normalization of the logarithm can in principle be also determined, by calculating finite parts in the integrals (\ref{SI}) and (\ref{SII}).

The circular Wilson loop at strong coupling thus has the following form:
\begin{equation}
 W(C_{\rm circle})=\,{\rm e}\,^{\sqrt{\lambda }A(a)},
\end{equation}
where
\begin{equation}
 A(a)=
\begin{cases}
 1+O(a^2) , & { a\rightarrow 0}
\\
 a+\frac{9\ln a}{32a}+O\left(\frac{1}{a}\right), & {a\rightarrow \infty }
\end{cases}
\end{equation}
To recover the usual units, $a$ should be replaced by $aM$.


\end{document}